\documentstyle[12pt]{article}
\begin{document}
\tolerance=5000
\def\pp{{\, \mid \hskip -1.5mm =}}
\def\cL{{\cal L}}
\def\be{\begin{equation}}
\def\ee{\end{equation}}
\def\bea{\begin{eqnarray}}
\def\eea{\end{eqnarray}}
\def\tr{{\rm tr}\, }
\def\nn{\nonumber \\}
\def\e{{\rm e}}
\def\D{{D \hskip -3mm /\,}}

\def\SEH{S_{\rm EH}}
\def\SGH{S_{\rm GH}}
\def\AdS5{{{\rm AdS}_5}}
\def\S4{{{\rm S}_4}}
\def\gfv{{g_{(5)}}}
\def\gfr{{g_{(4)}}}
\def\SC{{S_{\rm C}}}
\def\RH{{R_{\rm H}}}

\def\wlBox{\mbox{
\raisebox{0.1cm}{$\widetilde{\mbox{\raisebox{-0.1cm}\fbox{\ }}}$}}}
\def\htBox{\mbox{
\raisebox{0.1cm}{$\hat{\mbox{\raisebox{-0.1cm}{$\Box$}}}$}}}


\  \hfill 
\begin{minipage}{3.5cm}
May 2001 \\
\end{minipage}

\vfill

\begin{center}
{\large\bf On the way to Brane New World\footnote{
Based on the lectures delivered by S.N. at the 9th Numazu Meeting, 
March 8-10, 2001, held at Numazu College of Technology and 
18th Santo Seminar April 21, 2001, at Osaka University.}}

\vfill

{\sc Shin'ichi NOJIRI}\footnote{email: nojiri@cc.nda.ac.jp} 
and {\sc Sergei D. ODINTSOV}$^{\spadesuit}$\footnote{
On leave from Tomsk State Pedagogical University, RUSSIA. \\
\ \hskip 1cm email: odintsov@ifug5.ugto.mx}, \\

\vfill

{\sl Department of Applied Physics \\
National Defence Academy, 
Hashirimizu Yokosuka 239, JAPAN}

\vfill

{\sl $\spadesuit$ 
Instituto de Fisica de la Universidad de 
Guanajuato \\
Apdo.Postal E-143, 37150 Leon, Gto., MEXICO}

\vfill


{\bf ABSTRACT}

\end{center}

{\small 
In this report we
 consider brane-world universe (New Brane World) where an arbitrary large 
$N$ quantum CFT exists on the domain wall. 
This corresponds to implementing of Randall-Sundrum 
compactification within the context of AdS/CFT correspondence. 
Using anomaly induced effective action for domain wall CFT, 
the possibility of self-consistent quantum creation of 4d 
de Sitter wall universe (inflation) is demonstrated.
In case of maximally SUSY Yang-Mills theory the exact correspondence 
with radius and effective tension found by Hawking-Hertog-Reall is 
obtained. 

We also  discuss the bosonic sector of 5d gauged 
supergravity with single scalar and taking the boundary action 
as predicted by supersymmetry and discuss the possibility 
to supersymmetrize dilatonic New Brane World.
It is demonstrated that for a number of superpotentials the flat 
SUSY dilatonic brane-world (with dynamically induced brane dilaton) 
or quantum-induced de Sitter dilatonic brane-world (not Anti-de 
Sitter one) where SUSY is broken by the quantum effects occurs. 
The analysis of graviton perturbations indicates that gravity is 
localized on such branes.

New Brane World is useful in the study of FRW dynamics and cosmological 
 entropy bounds.
 Brane stress tensor is induced 
by quantum effects of dual CFT and brane crosses the horizon 
of AdS black hole.
The similarity between CFT entropy at the horizon and FRW equations 
is extended on the quantum level. This suggests the way to 
understand cosmological entropy bounds in quantum gravity.
}


\newpage

\section{Introduction}

The word ``Brane New World'' is the title of the paper 
by Hawking-Hertog-Reall \cite{HHR}.
After the discovery that gravity on the brane may be 
localized \cite{RS2}, there was renewed interest in the 
studies of higher-dimensional (brane-world) theories. 
In particular, numerous works \cite{CH} (and refs. therein) 
have been devoted to the investigation of cosmology 
(inflation) of brane-worlds. In refs.\cite{HHR,NOZ,NOplb} it 
has been suggested that the inflationary brane-world 
scenario could be realized due to quantum effects of 
brane matter. Such a scenario is based on the large $N$ 
quantum CFT living on the brane \cite{HHR,NOZ,NOplb}. 
Unlike to convenient brane-worlds, the boundary action is not 
the arbitrary one (the brane tension is not a free parameter). 
On the contrary, the surface terms on AdS-like space 
are motivated by the AdS/CFT correspondence. Their role is in 
making of variational procedure to be well-defined and in 
the elimination of the leading divergence of the AdS-like action. 
In accordance with AdS/CFT correspondence, there is quantum CFT 
living on the brane. Such brane quantum CFT produces conformal 
anomaly which leads to creation of effective brane tension. 
As a result, the dynamical mechanism to get flat or curved 
(de Sitter or Anti-de Sitter) brane-world 
appears \cite{HHR,NOZ,NOplb} in frames of AdS/CFT duality. 
Hence, one gets less fine-tuning in realization of brane-worlds as 
brane tension is not free parameter. The nice feature of 
this dynamical scenario is that the sign of conformal anomaly terms 
for usual matter predicts de Sitter (inflationary) universe as a 
preferrable solution in one-brane case. 

 From another side, there is much activity now in the 
supersymmetrization of Randall-Sundrum brane 
world \cite{susy1,susy2,BD,CLP} (see also refs. therein).
The 5d gauged supergravity represents very interesting model 
where supersymmetric dilatonic brane-world should be 
searched. Moreover, in such model, it is natural to try to 
construct supersymmetric dilatonic brane-world consistent 
with AdS/CFT correspondence \cite{AdS}. It could be then 
that such a scenario should be realized as supersymmetric 
version of New Brane World \cite{HHR,NOZ,NOplb,NOO}. 

Brane New World may find further applications. Indeed, it 
 is quite well-known fact that holographic principle suggests 
the interesting bounds between microscopic and Bekenstein-Hawking 
entropy \cite{H} as it was discussed in refs.\cite{HS,HMS}.
Recently, the very interesting attempt to study the holographic 
principle in Friedmann-Robertson-Walker (FRW) universe filled by 
CFT has been done by Verlinde \cite{EV}. Using dual 
AdS-description \cite{AdS} it has been found the relation 
between the entropy (energy) of CFT and the cosmological 
equations of motion in FRW universe. In particulary, the equation 
controlling the entropy bounds during evolution has been 
obtained \cite{EV} and Cardy-Verlinde formula has 
been derived. These results have been subsequently 
generalized and discussed in a number of works \cite{related,NO}.

One interesting extension has been presented in ref.\cite{SV} 
where similar questions have been studied from classical 
brane-world perspective\cite{RS2,RS1}. In particulary, the behaviour of
the CFT entropy at the horizon of bulk 5d AdS BH has been 
investigated and its comparison with FRW equations has 
been done. In the present report, based on \cite{NOent}, we 
also generalize the situation described in ref.\cite{SV} to 
the case of quantum-induced (or AdS/CFT induced) brane-worlds 
suggested in refs.\cite{HHR,NOZ,NOplb}. In this way, from one 
side, one gets quantum-corrected FRW universe equations as 
they look from the point of view of not only brane observer 
(who knows nothing about bulk 5d BH) but also from the point of 
view of quantum induced brane-world. From another side, one 
gets the quantum-corrected brane entropy as well as Hubble 
constant and Hawking temperature at the horizon. Finally, 
this may be considered as extension of scenario of 
refs.\cite{HHR,NOZ,NOplb} (see refs.\cite{extension} for related 
questions) which admits also generalization for the presence 
of non-trivial dilaton and (or) supersymmetrization \cite{NOO}  
for the case when brane crosses the horizon of AdS-black hole.

This report is organized as follows. In the next section, 
the equivalence between 5d dilatonic gravity and 4d dilatonic 
gravity coupled with CFT is discussed. In section 3, based 
on \cite{NOplb}, we give the inflationary brane-world scenario 
realized due to quantum effects of brane matter by using anomaly 
induced effective action. As a result, one can consider 
the arbitrary content of CFT living on the wall. Moreover, the 
formalism is applied not only to 4d de Sitter wall but also to 4d
hyperbolic wall in 5d Anti-de Sitter universe or 4d conformally 
flat universe. In section 4, the review of the construction of 
classical supersymmetric brane-world is done for bosonic sector 
of 5d gauged SG with single dilaton. Boundary action is 
predicted by supersymmetry. Half of supersymmetries survives 
for flat brane-world (as it follows from the analysis of 
BPS condition). The classical SUSY curved brane-worlds cannot 
be realized. 
Fifth section is devoted to the extension of 
the analysis of fourth section modified by the quantum 
contribution from brane CFT in order to construct SUSY 
New Brane-World. It is shown for number of superpotentials that, 
unlike to classical case, the quantum induced de Sitter 
brane-world is created. However, the brane supersymmetry is 
broken by quantum effects. The example of SUSY flat brane-world, 
where boundary value of dilaton is defined by quantum effects, 
is also given. In section 6, the analysis of graviton 
perturbations around found solutions is done. It is shown 
that only one normalizable solution corresponding to zero 
mode exists. In other words, gravity should be trapped on 
the brane in such scenario.
In section 7, we consider the generalization of approach of ref.\cite{SV} 
to the case of quantum-induced brane-worlds  
\cite{HHR,NOZ,NOplb} and obtain a quantum-corrected FRW universe 
equations. Quantum-corrected Hubble constant, Hawking 
temperature and cosmological entropy are found on the FRW brane. 
Some brief summary and outlook is given in final section.

\section{AdS/CFT and the localization of the gravity}

AdS$_5$/CFT$_4$ correspondence tells us that the effective 
action $W_{\rm CFT}$ of CFT in 4 dimensions is given by the path 
integral of the supergravity in 5 dimensional AdS space:
\bea
\label{A1}
&& \e^{-W_{\rm CFT}}=\int [dg][d\varphi]\e^{-S_{\rm grav}}\ ,\quad 
S_{\rm grav}=\SEH + \SGH + S_1 + S_2 + \cdots\ , \\
&& \SEH={1 \over 16\pi G}\int d^5 x \sqrt{\gfv}\left(R_{(5)} 
 + {12 \over l^2} + \cdots \right)\ ,\quad 
\SGH={1 \over 8\pi G}\int d^4 x \sqrt{\gfr}\nabla_\mu n^\mu\ , \nn
&&S_1= -{1 \over 8\pi G}\int d^4 x \sqrt{\gfr}\left({3 \over l}
+ \cdots \right)\ ,\quad 
S_2= -{l \over 16\pi G}\int d^4 x \sqrt{\gfr}\left(
{1 \over 2}R_{(4)} + \cdots \right)\ .\nonumber 
\eea
Here $\varphi$ expresses the (matter) fields besides the graviton. 
$\SEH$ corresponds to the Einstein-Hilbert action and 
$\SGH$ to the Gibbons-Hawking surface counter term and 
$n^\mu$ is the unit vector normal to the boundary. 
$S_1$, $S_2$, $\cdots$ correspond to the surface counter terms, 
which cancell the divergences when the boundary in AdS$_5$ goes to 
the infinity.\footnote{See \cite{NOOg} for the surface 
 counterterms  in dilaton coupled supergravities.} 

In \cite{HHR}, two 5 dimensional balls B$_5^{(1,2)}$ are 
glued on the boundary, which is 4 dimensional sphere 
S$_4$\footnote{The reason why this situation was considered 
is given in the next section. In this section, we can consider 
the case that the brane is the boundary of two AdS spaces.}. 
Instead of $S_{\rm grav}$, if one considers the following 
action $S$
\be
\label{A2}
S=\SEH + \SGH + S_1=S_{\rm grav} - S_2 - \cdots,
\ee
for two balls, using (\ref{A1}), one gets the following 
boundary theory in terms of the partition function \cite{HHR}:
\bea
\label{A3}
&& \int_{{\rm B}_5^{(1)} + {\rm B}_5^{(1)} 
+{\rm S}_4} [dg][d\varphi]\e^{-S} 
= \left(\int_{{\rm B}_5} [dg][d\varphi]\e^{-\SEH - \SGH - S_1} 
\right)^2 \nn
&& =\e^{2S_2 + \cdots}\left(\int_{{\rm B}_5} [dg][d\varphi]
\e^{-S_{\rm grav}} \right)^2 =\e^{-2W_{\rm CFT}+2S_2 + \cdots}\ .
\eea
Since $S_2$ can be regarded as the Einstein-Hilbert action on 
the boundary, which is S$_4$ in the present case, the gravity 
on the boundary becomes dynamical. The 4 dimensional gravity 
is nothing but the gravity localized on the brane in the 
Randall-Sundrum model \cite{RS2}. 

For ${\cal N}=4$ $SU(N)$ Yang-Mills theory, the AdS/CFT dual is 
given by identifying
\be
\label{AdSCFT}
l=g_{\rm YM}^{1 \over 2}N^{1 \over 4}l_s\ ,
\quad {l^3 \over G}={2N^2 \over \pi}\ .
\ee
Here $g_{\rm YM}$ is the coupling of the Yang-Mills theory and 
$l_s$ is the string length. Then (\ref{A3}) tells that the 
RS model is equivalent to  a CFT (${\cal N}=4$ $SU(N)$ 
Yang-Mills theory) coupled to 4 dimensional gravity including 
some correction coming from the higher order counter terms with 
a Newton constant given by $G_4=G/l$. 
This is an excellent explanation \cite{HHR} to why gravity 
is trapped on the brane.

\section{Brane New World}

In \cite{RS2}, the discussion was limited by the flat 
brane. In this case, however, the brane crosses the event horizon 
in the finite time, which opens the causality problem\footnote{
We have the following causal structure.

\setlength{\unitlength}{0.5mm}
\begin{picture}(200,100)
\put(20,5){\line(0,1){90}}
\put(10,0){spacial infinity}
\put(25,82){future}
\put(23,77){horizon}
\put(25,20){past}
\put(23,15){horizon}
\put(55,45){brane}
\thicklines
\put(20,50){\line(1,1){45}}
\put(20,50){\line(1,-1){45}}
\qbezier[1000](65,95)(30,50)(65,5)
\end{picture}
}. 
To avoid this problem, it would be natural to consider de Sitter 
brane which is also motivated by cosmology\cite{HHR}. 
If the brane is de Sitter space, the brane does not 
cross the horizon. 
Motivated by this, we consider the curved brane in this 
section.

 Let us take the spacetime whose boundary is 4 dimensional 
sphere $\S4$, which can be identified with a D3-brane. 
The bulk part is given by 5 dimensional 
Euclidean anti de Sitter space $\AdS5$\footnote{  For the 
expressions of the metric of AdS, see Appendix.}
\be
\label{AdS5i}
ds^2_\AdS5=dy^2 + l^2 \sinh^2 {y \over l}d\Omega^2_4\ .
\ee
Here $d\Omega^2_4$ is given by the metric of $\S4$ 
with unit radius. One also assumes the boundary (brane) 
lies at $y=y_0$ 
and the bulk space is given by gluing two regions 
given by $0\leq y < y_0$. 

We start with the action $S$ which is the sum of 
the Einstein-Hilbert action $\SEH$, the Gibbons-Hawking 
surface term $\SGH$,  the surface counter term $S_1$ 
and the trace anomaly induced action 
${\cal W}$\footnote{For the introduction to
anomaly induced effective action in curved space-time (with torsion), see
section 5.5 in \cite{BOS}.}: 
\bea
\label{Stotal}
&& S=\SEH + \SGH + 2 S_1 + {\cal W}\ ,\quad 
\SEH={1 \over 16\pi G}\int d^5 x \sqrt{\gfv}\left(R_{(5)} 
+ {12 \over l^2}\right)\ , \nn
&& \SGH={1 \over 8\pi G}\int d^4 x \sqrt{\gfr}\nabla_\mu n^\mu 
\ ,\quad S_1= -{3 \over 8\pi Gl}\int d^4 x \sqrt{\gfr} \ ,\nn
&&{\cal W}= b \int d^4x \sqrt{\widetilde g}\widetilde F A 
 + b' \int d^4x \sqrt{\widetilde g}
\left\{A \left[2{\widetilde\Box}^2 
+\widetilde R_{\mu\nu}\widetilde\nabla_\mu\widetilde\nabla_\nu 
 - {4 \over 3}\widetilde R \widetilde\Box^2 \right.\right. \nn
&& \left.\left. \qquad 
+ {2 \over 3}(\widetilde\nabla^\mu \widetilde R)\widetilde\nabla_\mu
\right]A 
+ \left(\widetilde G - {2 \over 3}\widetilde\Box \widetilde R
\right)A \right\} \nn
&& \qquad -{1 \over 12}\left\{b''+ {2 \over 3}(b + b')\right\}
\int d^4x \sqrt{\widetilde g} \left[ \widetilde R - 6\widetilde\Box A 
 - 6 (\widetilde\nabla_\mu A)(\widetilde \nabla^\mu A)
\right]^2 \ .
\eea 
Here the quantities in the  5 dimensional bulk spacetime are 
specified by the suffices $_{(5)}$ and those in the boundary 4 
dimensional spacetime are by $_{(4)}$. 
The factor $2$ in front of $S_1$ in (\ref{Stotal}) is coming from 
that we have two bulk regions which 
are connected with each other by the brane. 
In (\ref{Stotal}), $n^\mu$ is 
the unit vector normal to the boundary. In (\ref{actions2}), one chooses 
the 4 dimensional boundary metric as 
\be
\label{tildeg}
\gfr_{\mu\nu}=\e^{2A}\tilde g_{\mu\nu}
\ee 
and we specify the 
quantities with $\tilde g_{\mu\nu}$ by using $\tilde{\ }$. 
$G$ ($\tilde G$) and $F$ ($\tilde F$) are the Gauss-Bonnet
invariant and the square of the Weyl tensor, which are given as
\footnote{We use the following curvature conventions:
\begin{eqnarray*}
&& R=g^{\mu\nu}R_{\mu\nu}\ ,\quad 
R_{\mu\nu}= R^\lambda_{\ \mu\lambda\nu} \\
&& R^\lambda_{\ \mu\rho\nu}=-\Gamma^\lambda_{\mu\rho,\nu}
+ \Gamma^\lambda_{\mu\nu,\rho}
- \Gamma^\eta_{\mu\rho}\Gamma^\lambda_{\nu\eta}
+ \Gamma^\eta_{\mu\nu}\Gamma^\lambda_{\rho\eta}\ ,\quad 
\Gamma^\eta_{\mu\lambda}={1 \over 2}g^{\eta\nu}\left(
g_{\mu\nu,\lambda} + g_{\lambda\nu,\mu} - g_{\mu\lambda,\nu} 
\right)\ .
\end{eqnarray*}}
\be
\label{GF}
G=R^2 -4 R_{ij}R^{ij}
+ R_{ijkl}R^{ijkl} \ ,\quad 
F={1 \over 3}R^2 -2 R_{ij}R^{ij}
+ R_{ijkl}R^{ijkl} \ ,
\ee
In the effective action (\ref{actions2}) induced by brane quantum 
matter, in general, with $N$ real scalar, $N_{1/2}$ 
Dirac spinor, $N_1$ vector 
fields, $N_2$  ($=0$ or $1$) gravitons and $N_{\rm HD}$ higher 
derivative conformal scalars, $b$, $b'$ and $b''$ are\footnote{ 
These parameters appear in the general expression of the 
conformal anomaly $T$
\[
T=b\left(F+{2 \over 3}\Box R\right) + b' G + b''\Box R\ .
\]
}
\bea
\label{bs}
&& b={N +6N_{1/2}+12N_1 + 611 N_2 - 8N_{\rm HD} 
\over 120(4\pi)^2}\ ,\nn 
&& b'=-{N+11N_{1/2}+62N_1 + 1411 N_2 -28 N_{\rm HD} 
\over 360(4\pi)^2}\ ,\quad b''=0\ .
\eea
Usually, $b''$ may be changed by the finite renormalization 
of local counterterm in the gravitational effective action. 
As it was the case in ref.\cite{NOO}, the term proportional 
to $\left\{b''+ {2 \over 3}(b + b')\right\}$ in (\ref{actions2}), and 
therefore $b''$ does not contribute to the equations of motion.

For typical examples motivated by AdS/CFT correspondence\cite{AdS} 
one has:
a) ${\cal N}=4$ $SU(N)$ SYM theory\footnote{
A multiplet of ${\cal N}=4$ theory contains 1 vector, 4 Majorana 
spinors (2 Dirac spinors) and 6 real scalars.
} $b=-b'={C \over 4}={N^2 -1 \over 4(4\pi )^2}$, 
b) ${\cal N}=2$ $Sp(N)$ theory\footnote{
${\cal N}=2$ $Sp(N)$ theory contains $2N^2 + N$ vector 
multiplets and $2N^2 + 7N -1$ hypermultiplets. 
One vector multiplet contains 1 vector, 2 Majorana spinors 
(1 Dirac spinor) and 2 real scalars. On the other hand, 
one hypermultiplet does not contain a vector but 2 Majorana 
spinors (1 Dirac spinor) and 4 real (2 complex) scalars.
} $b={12 N^2 + 18 N -2 \over 24(4\pi)^2}$, 
$b'=-{12 N^2 + 12 N -1 \over 24(4\pi)^2}$. 
 Note that $b'$ is negative in the above cases.

We should also note that ${\cal W}$ in (\ref{actions2}) is defined up to 
conformally invariant functional, which cannot be determined 
from only the conformal anomaly. The conformally flat space is a pleasant 
exclusion where anomaly induced effective action is defined uniquely.
However, one can argue that such conformally invariant functional gives 
next to leading contribution as mass parameter of regularization 
may be adjusted to be arbitrary small (or large).

The metric of $\S4$ with the unit radius is given by
\be
\label{S4metric1}
d\Omega^2_4= d \chi^2 + \sin^2 \chi d\Omega^2_3\ .
\ee
Here $d\Omega^2_3$ is described by the metric of 3 dimensional 
unit sphere. If we change the coordinate $\chi$ to 
$\sigma$ by $\sin\chi = \pm {1 \over \cosh \sigma}$, 
one obtains\footnote{
If we Wick-rotate the metric by $\sigma\rightarrow it$, we 
obtain the metric of de Sitter space:
\[
d\Omega^2_4\rightarrow ds_{\rm dS}^2
= {1 \over \cos^2 t}\left(-dt^2 + d\Omega^2_3\right)\ .
\]
}
\be
\label{S4metric2}
d\Omega^2_4= {1 \over \cosh^2 \sigma}\left(d \sigma^2 
+ d\Omega^2_3\right)\ .
\ee
On the other hand, the metric of the 4 dimensional flat 
Euclidean space is given by
\be
\label{E4metric}
ds_{\rm 4E}^2= d\rho^2 + \rho^2 d\Omega^2_3\ .
\ee
Then by changing the coordinate as $\rho=\e^\sigma$, one gets
\be
\label{E4metric2}
ds_{\rm 4E}^2= \e^{2\sigma}\left(d\sigma^2 + d\Omega^2_3\right)\ .
\ee
For the 4 dimensional hyperboloid with the unit radius, 
the metric is given by
\be
\label{H4metric1}
ds_{\rm H4}^2= d \chi^2 + \sinh^2 \chi d\Omega^2_3\ .
\ee
Changing the coordinate $\chi$ to $\sigma$ by 
$\sinh\chi = {1 \over \sinh \sigma}$, 
one finds
\be
\label{H4metric2}
ds_{\rm H4}^2 = {1 \over \sinh^2 \sigma}\left(d \sigma^2 
+ d\Omega^2_3\right)\ .
\ee

 Let us now discuss the 4 dimensional hyperboloid whose boundary 
is the 3 dimensional sphere $S_3$ but we can consider the cases that 
the boundary is a 3 dimensional flat Euclidean space $R_3$ or a 
3 dimensional hyperboloid $H_3$. We will, however, only consider 
the case that the boundary is $S_3$ since the results for other 
cases are almost equivalent.

Motivated by (\ref{AdS5i}), (\ref{S4metric2}), 
(\ref{E4metric2}) and (\ref{H4metric2}), one assumes 
the metric of 5 dimensional space time as follows:
\be
\label{metric1}
ds^2=dy^2 + \e^{2A(y,\sigma)}\tilde g_{\mu\nu}dx^\mu dx^\nu\ ,
\quad \tilde g_{\mu\nu}dx^\mu dx^\nu\equiv l^2\left(d \sigma^2 
+ d\Omega^2_3\right)
\ee
and one identifies $A$ and $\tilde g$ in (\ref{metric1}) with those in 
(\ref{tildeg}). Then $\tilde F=\tilde G=0$, 
$\tilde R={6 \over l^2}$ etc. 
Due to the assumption (\ref{metric1}), the actions in (\ref{Stotal}) 
have the following forms:
\bea
\label{actions2}
&& \SEH= {l^4 V_3 \over 16\pi G}\int dy d\sigma \left\{\left( -8 
\partial_y^2 A - 20 (\partial_y A)^2\right)\e^{4A} \right. \nn
&& \qquad \left. +\left(-6\partial_\sigma^2 A 
 - 6 (\partial_\sigma A)^2 
+ 6 \right)\e^{2A} + {12 \over l^2} \e^{4A}\right\} \nn
&& \SGH= {l^4 V_3 \over 2\pi G}\int d\sigma \e^{4A} 
\partial_y A \ ,\quad 
S_1= - {3l^3 V_3 \over 8\pi G}\int d\sigma \e^{4A} \nn
&& {\cal W}= V_3 \int d\sigma \left[b'A\left(2\partial_\sigma^4 A
 - 8 \partial_\sigma^2 A \right) 
 - 2(b + b')\left(1 - \partial_\sigma^2 A 
 - (\partial_\sigma A)^2 \right)^2 \right]\ .
\eea
Here $V_3$ is the volume or area of the unit 3 sphere. 

In the bulk, one obtains the following equation of motion 
from $\SEH$ by the variation over $A$:
\be
\label{eq1}
0= \left(-24 \partial_y^2 A - 48 (\partial_y A)^2 
+ {48 \over l^2}
\right)\e^{4A} + {1 \over l^2}\left(-12 \partial_\sigma^2 A 
- 12 (\partial_\sigma A)^2 + 12\right)\e^{2A}\ ,
\ee
which corresponds to one of the Einstein equations. 
Then one finds solutions, $A_S$, which corresponds to 
the metric of $\AdS5$ in (\ref{AdS5i}) with (\ref{S4metric2}), 
$A_E$, which corresponds to (\ref{E4metric2}), and $A_H$, which 
corresponds to (\ref{H4metric2}). 
\bea
\label{blksl}
&& A=A_S,\ A_E,\ A_H, \nn
&& A_S=\ln\sinh{y \over l} - \ln \cosh\sigma\ ,\quad 
A_E={y \over l} + \sigma\ ,\quad 
A_H=\ln\cosh{y \over l} - \ln\sinh\sigma\ .
\eea
One should note that all the metrics in (\ref{blksl}) 
locally describe the 
same spacetime, that is the local region of $\AdS5$, in 
the bulk. As we assume, however, that there is a brane at 
$y=y_0$, the shapes of the branes are different from each 
other due to the choice of the metric. 

On the brane at the boundary, 
one gets the following equation:\footnote{
We should note that the radial ($y$) component of the 
geodesic equation for the  in the metric (\ref{metric1}) 
is given by ${d^2x^y \over d\tau^2} + \partial_y A \e^{2A}
\left({dx^t \over d\tau}\right)^2=0$. Here $\tau$ is the 
proper time and one can normalize $\e^{2A}
\left({dx^t \over d\tau}\right)^2 = 1$ and obtain 
${d^2x^y \over d\tau^2} + \partial_y A=0$. 
Therefore the classical part in (\ref{eq2}) expresses 
the balance of the gravitational force and tension. 
 The mass density of the brane is 
given by ${3 \over 8\pi G}$.} 
\bea
\label{eq2}
0&=&{48 l^4 \over 16\pi G}\left(\partial_y A - {1 \over l}
\right)\e^{4A}
+b'\left(4\partial_\sigma^4 A - 16 \partial_\sigma^2 A\right) \nn
&& - 4(b+b')\left(\partial_\sigma^4 A + 2 \partial_\sigma^2 A 
 - 6 (\partial_\sigma A)^2\partial_\sigma^2 A \right)\ .
\eea
We should note that the contributions from $\SEH$ and $\SGH$ are 
twice from the naive values since we have two bulk regions which 
are connected with each other by the brane. 
Substituting the bulk solution $A=A_S$ in (\ref{blksl}) into 
(\ref{eq2}) and defining the radius $R$ of the brane by
$R\equiv l\sinh{y_0 \over l}$, one obtains
\be
\label{slbr2}
0={1 \over \pi G}\left({1 \over R}\sqrt{1 + {R^2 \over l^2}}
 - {1 \over l}\right)R^4 + 8b'\ .
\ee
Note that eq.(\ref{slbr2}) does not depend on $b$. 
This equation generalizes the corresponding result of ref.\cite{HHR} 
for the case when the arbitrary amount of quantum conformal 
matter sits on de Sitter wall. Adopting AdS/CFT 
correspondence one can argue that in symmetric phase the 
quantum brane matter comes due to maximally SUSY Yang-Mills theory.

As we have $b'\rightarrow -{N^2 \over 4(4\pi )^2}$ 
in case of the large $N$ limit of ${\cal N}=4$ $SU(N)$ SYM theory, 
we find
\be
\label{slbr3}
{R^3 \over l^3}\sqrt{1 + {R^2 \over l^2}}={R^4 \over l^4}
+ {GN^2 \over 8\pi l^3}\ ,
\ee 
which exactly coincides with the result in \cite{HHR}. 
This equation has the unique solution for positive radius 
which defines brane-world de Sitter universe (inflation) 
induced by quantum effects.

On the other hand, if one substitutes the solution $A=A_E$ in 
(\ref{blksl}), corresponding to flat Euclidean brane 
into (\ref{eq2}), we find that (\ref{eq2}) is always 
(independent of $y_0$) satisfied since 
$\partial_y A={1 \over l}$ and $\partial_\sigma^2 A=0$. 

If one substitutes $A=A_H$ in (\ref{blksl}), which corresponds 
to the brane with the shape of the hyperboloid, and one 
defines the radius $\RH$ of the brane by 
$\RH\equiv l\cosh{y_0 \over l}$, then 
\be
\label{slbr2b}
0={1 \over \pi G}\left(\pm{1 \over \RH}\sqrt{-1 
+ {\RH^2 \over l^2}} - {1 \over l}\right)\RH^4 + 8b'\ .
\ee
We should note that eq.(\ref{slbr2b}) does not depend on $b$ 
again. In order that Eq.(\ref{slbr2b}) has a solution, $b'$ 
must be positive, which conflicts with the case of 
${\cal N}=4$ $SU(N)$ SYM theory or usual conformal matter. 
In general, however,  for some exotic theories, like
higher derivative conformal scalar\footnote{Such higher derivative 
conformal scalar naturally appears in infra-red sector of 
quantum gravity\cite{AMO}.}, $b'$ can be positive and one can assume 
for the moment that $b'>0$ here. 
Hence, we showed that quantum, conformally invariant matter 
on the wall, leads to the inducing of inflationary 4d de 
Sitter-brane universe realized within 5d Anti-de Sitter space 
(a la Randall-Sundrum\cite{RS1,RS2}). Of course, analytical continuation 
of our 4d sphere to Lorentzian signature is supposed what leads 
to ever expanding inflationary brane-world universe. 
In 4d QFT (no higher dimensions) such idea of anomaly 
induced inflation  has been suggested long ago in refs.\cite{SMM}.
On the same time the inducing of 4d hyperbolic wall in brane universe is
highly suppressed
and may be realized only for exotic conformal matter. 
The analysis of the role of domain wall CFT to metric 
fluctuations may be taken from results of ref.\cite{HHR}. 

It is interesting to note that our approach is quite general. In 
particulary, it is not difficult to take into account the quantum gravity 
effects (at least, on the domain wall). That can be done by using the 
corresponding analogs of central charge for various QGs which 
may be taken 
from beta-functions listed in book \cite{BOS}. In other words, there will 
be only QG contributions to coefficients $b,b'$ but no more changes in
subsequent equations. 

Generalizations to the cases of the gravity coupled with 
dilaton are given in \cite{NOO,NOOs}. Especially in \cite{NOOs}, 
supersymmetric cases are discussed. This will be considered below.

\section{Classical supersymmetric brane-world} 

The 5d ${\cal N}=8$ gauged supergravity can be obtained from 10d IIB 
supergravity, where the spacetime is compactified into 
S$_5\times$M$_5$. Here S$_5$ is 5d sphere and M$_5$ 
is a 5d manifold, where gauged supergravity lives. 
The bosonic sector (gravity and scalar part) of the 5d 
gauged supergravity is given by
\be
\label{Sbulk}
S_{\rm bulk}={1 \over 16\pi G}\int d^5 x \sqrt{-\gfv}\left(R_{(5)} 
 -{1 \over 2}g_{ij}(\phi_k)\nabla_\mu\phi^i\nabla^\mu \phi^j 
 - V(\phi^i) \right) \ .
\ee
In the 5-dimensional maximal supergravity, the scalar field 
parametrizes the coset of $E_6/SL(6,R)$. In (\ref{Sbulk}), 
$g_{ij}(\phi^k)$ is the induced scalars metric and the 
potential $V(\phi^i)$ can be rewritten in terms of the 
superpotential $W(\phi_i)$:\footnote{Eq.(\ref{SuW0}) 
can be regarded as the definition of $W(\phi_i)$ for  
rather general potential $V(\phi)$ even if there is no 
supersymmetry. The corresponding potential in case of
 $D=5$ ${\cal N}=8$ supergravity, 
including higher rank tensors, was found in \cite{GRW}.
The potential under discussion may be considered as the one 
corresponding to some its subsector, for example, as the one discussed 
in last of refs.\cite{CLP}. 
}
\be
\label{SuW0}
V(\phi_i)={3 \over 4}\left({3 \over 2}
g^{ij}(\phi_k){dW(\phi_k) \over d\phi^i}{dW(\phi_k) \over d\phi^j}
 - W(\phi_k)^2\right)\ .
\ee
We choose the boundary action $S_{\rm bndry}$ in 
the following form:
\be
\label{Sbndry2}
S_{\rm bndry}=\mp {3 \over 16\pi G} \int d^4 x 
\sqrt{-\gfr}W(\phi)\ .
\ee
This tells that the brane is BPS saturated state, that is, the 
brane preserves the half of the supersymmetries of the whole 
system. 
In order to see it, one considers the simple case that only one 
scalar field $\phi$ is non-trivial and $g_{\phi\phi}=1$ 
and investigate the equations of motion given by the simplified 
action: 
\be
\label{S}
S= {1 \over 16\pi G}\left[ \int_M d^5 x
 \sqrt{-\gfv}\left( R_{(5)}
 - {1 \over 2}\partial_\mu\phi \partial^\mu\phi
 - V(\phi) \right) - \sum_{i=1,2}\int_{B_i} d^4 x \sqrt{-\gfr} 
U_i(\phi)\right]\ .
\ee
Here $B_i$'s express the boundaries or branes. 
This model generalizes that in \cite{RS1}, where two branes 
are boundaries of the bulk spacetime. The model with only one 
brane can be obtained, of course by letting one of the 
branes go to infinity. 
At first, we do not specify the form of $U_i(\phi)$ but by 
investigating the equations of motion, we will see the 
correspondence with (\ref{Sbndry2}). 
 Assume the metric in 5d spacetime as 
\be
\label{Mi}
ds^2=dz^2 + \e^{2A(z)}\eta_{ij}dx^i dx^j\ ,
\ee
and $\phi$ only depends on $z$. One also supposes the branes 
sit on $z=z_1$ and $z=z_2$, respectively.
Then the equations of motion are given by
\bea
\label{Ei}
\phi''+ 4A'\phi' &=& {d V \over d \phi} + \sum_{i=1,2}
{d U_i(\phi) \over d \phi} \delta(z-z_i)\ , \\
\label{Eii}
4A''+ 4(A')^2 + {1 \over 2}(\phi')^2 
&=& -{V \over 3}  - {2 \over 3}\sum_{i=1,2}
U_i(\phi)\delta(z-z_i) \ , \\
\label{Eiii}
A'' + 4 (A')^2 &=& -{V \over 3}
 - {1 \over 6}\sum_{i=1,2}U_i(\phi)\delta(z-z_i) \ .
\eea
Here $'\equiv {d \over dz}$.
For purely bulk sector ($z_1<z<z_2$, as we assume $z_1<z_2$), 
Eqs. (\ref{Ei}-\ref{Eiii}) have the following first integrals:
\bea
\label{Iii}
\phi'={3 \over 2}{d W \over d\phi}\ ,
\quad A' = - {1 \over 4}W\ .
\eea
(Here the ambiguity in the sign when solving Eq.(\ref{SuW0}) 
with respect to $W(\phi)$ is absorbed into the definition 
of $W(\phi)$.)  One should note that classical solutions do not 
always satisfy the above Eqs. in (\ref{Iii}). Classical 
solutions are generally not invariant under the supersymmetry 
transformations and the supersymmetry in the bulk is broken 
in the classical background. Eq.(\ref{Iii}) is nothing but the 
condition that the classical solution is invariant under the 
half of the supersymmetry transformations. When there are 
branes, any solution of the equations of motion including 
the equation coming from the branes might not satisfy 
Eqs.(\ref{Iii}). We now investigate the condition for the 
brane action which allows solution satisfying Eqs.(\ref{Iii}).
Then some of the supersymmetries are preserved in the whole system. 

Near the branes, Eqs.(\ref{Ei}-\ref{Eiii}) have the forms  
$\phi'' \sim {d U_i(\phi)\over d\phi}\delta (z-z_i)$, 
$A'' \sim -{U_i(\phi) \over 6}\delta (z-z_i)$. Then
\be
\label{Eivb}
2\phi' \sim {d U_i(\phi)\over d\phi}{\rm sgn}(z-z_i)\ ,
\quad 2A' \sim -{U_i(\phi) \over 6}{\rm sgn}(z-z_i) \ ,
\ee
at $z=z_i$.\footnote{
Here the function ${\rm sgn}(x)$ is defined by
\[
{\rm sgn}(x)=\left\{
\begin{array}{ll}
1 \ & x>0 \\
-1 & x<0 \\
\end{array}
\right.
\]
and $z$ is the coordinate perpendicular to the boundary or 
brane.} 
If Eqs.(\ref{Iii}) are satisfied, one finds
\be
\label{Ev}
U_1(\phi)= 3W(\phi)\ ,\quad
U_2(\phi)=- 3W(\phi)\ .
\ee
Eq.(\ref{Ev}) reproduces Eq.(\ref{Sbndry2}). 
Note that Eq.(\ref{Iii}) is nothing but the BPS condition, 
where the half of the supersymmetries in the whole system 
are preserved. As we are considering the solution where 
fermionic fields vanish, the variations of the fermionic fields 
under the supersymmetry transformation should vanish if the solution 
preserves the supersymmetry. If Eq.(\ref{Iii}) is satisfied, 
the variations of gravitino and dilatino vanish under the half 
of the supersymmetry transformation.  

As an extension, one can consider the case that the brane is
curved. Instead of (\ref{Mi}), we take the following metric:
\be
\label{Mi2}
ds^2=dz^2 + \e^{2A(z)}\tilde g_{ij}dx^i dx^j\ ,
\ee
Here $\tilde g_{ij}$ is the metric of the Einstein manifold,
which is defined by
\be
\label{Ein}
\tilde R_{ij}=k \tilde g_{ij}\ ,
\ee
where $\tilde R_{ij}$ is the Ricci tensor given by
$\tilde g_{ij}$ and $k$ is a constant.
Then Eqs.(\ref{Ei}) and (\ref{Eii}) do no change but
one obtains the following equation instead of (\ref{Eiii}):
\be
\label{Eiiib}
A'' + 4 (A')^2 = k\e^{2A} - {V \over 3} 
 - {1 \over 6}\sum_{i=1,2}
U_i(\phi)\delta(z-z_i) \ .
\ee
Especialy when $k=0$,  one gets the previous solution for $\phi$,
$A$ and $U_i$. 
Even for $k=0$, the brane is not always flat, for example, if 
as $\tilde g_{ij}$ in (\ref{Mi2}), we choose the metric of the 
Schwarzschild black hole or Kerr black hole spacetime, then 
Eq.(\ref{Ein}) is satisfied since the Ricci tensor vanishes. 
 
Therefore the brane solutions with these black holes of $k=0$ 
would preserve the supersymmetry of the whole system. 
When $k\neq 0$, however, one finds that Eq.(\ref{Eiiib}) has no 
solution which satisfies the BPS condition (\ref{Iii}). 
This might tell that classical curved brane breaks the 
supersymmetry in such formalism. 
When $k>0$, the brane is 4d de Sitter space or 4d sphere when 
Wick-rotated to the Euclidean signature. On the other hand, when 
$k<0$, the brane is 4d anti-de Sitter space or 4d hyperboloid in 
the Euclidean signature.

\section{Supersymmetric brane new world}

If 10d spacetime, where IIB supergraviry lives, is 
compactified into S$_5\times$M$_5$, we effectively obtain 
5d ${\cal N}=8$ gauged supergravity in the bulk and 
4d ${\cal N}=4$ $SU(N)$ or $U(N)$ super-Yang-Mills theory 
coupled with (super)gravity on the brane. On the other hand, 
if 10d spacetime is compactified into X$_5\times $M$_5$, where 
X$_5$ is S$_5/Z_2$, ${\cal N}=2$ $Sp(N)$ super-Yang-Mills 
theory coupled with (super)gravity would be realized on the 
brane. Since the matter multiplets of the super-Yang-Mills are 
coupled with (super)gravity, they generate a conformal 
anomaly on quantum level. 
 
In this section, we include the trace anomaly induced 
action on the brane to the analysis of supersymmetric 
brane-world. One chooses the brane action to preserve 
the supersymmetry as in the previous section and consider the 
solution where the scalar field is non-trivial. 
Here  mainly Euclidean signature is used. 

As curved brane is considered, we assume that 
the metric of (Euclidean) AdS has the following form:
\be
\label{AdS}
ds^2=dz^2 + \sum_{i,j=1}^4 g_{(4)ij} dx^i dx^j\ ,
\quad g_{(4)ij}=\e^{2\tilde A(z)}\hat g_{ij}\ .
\ee
Here $\hat g_{ij}$ is the metric of the Einstein manifold as in 
(\ref{Ein}). 
One can consider two copies of the regions given by $z<z_0$ and 
glue two regions putting a brane at $z=z_0$. 

Let us start with Euclidean signature action $S$ which is 
 sum of the Einstein-Hilbert action $\SEH$ with kinetic term 
and potential $V(\phi)$ for dilaton $\phi$, the Gibbons-Hawking 
surface term $\SGH$, the surface counterterm $S_1$ and the 
trace anomaly induced action ${\cal W}$ \cite{NOO}: 
\bea
\label{Stotalb}
&& S=\SEH^\phi + \SGH + 2 S_1^\phi + \tilde{\cal W}, \nn 
&& \SEH^\phi={1 \over 16\pi G}\int d^5 x \sqrt{\gfv}\left(R_{(5)} 
 -{1 \over 2}\nabla_\mu\phi\nabla^\mu \phi 
 - V(\phi) \right), \nn 
&& \SGH={1 \over 8\pi G}\int d^4 x \sqrt{\gfr}\nabla_\mu n^\mu\ , 
\quad S_1^\phi= -{3 \over 16\pi G l}\int d^4 x 
\sqrt{\gfr} W(\phi), \nn
&& {\cal W}^\phi= {\cal W} + C \int d^4x \sqrt{\widetilde g}
A \phi \left[{\wlBox}^2 
+ 2\widetilde R_{\mu\nu}\widetilde\nabla_\mu\widetilde\nabla_\nu 
 - {2 \over 3}\widetilde R \wlBox^2 
+ {1 \over 3}(\widetilde\nabla^\mu \widetilde R)\widetilde\nabla_\mu
\right]\phi \ .
\eea 
Here the quantities in the  5 dimensional bulk spacetime are 
specified by the suffices $_{(5)}$ and those in the boundary 4 
dimensional spacetime are specified by $_{(4)}$. 
${\cal W}$ in (\ref{Stotalb}) is defined in (\ref{actions2}). 

In \cite{NOO}, as an action on the brane, corresponding to 
$S_1^\phi$ in (\ref{Stotalb}), the action motivated by the 
counterterm method in AdS/CFT correspondence was used:
\be
\label{S1noo}
S_1^{\rm NOO}= -{1 \over 16\pi G l}\int d^4 x \sqrt{\gfr}\left(
{6 \over l} + {l \over 4}\Phi(\phi)\right)\ .
\ee
In the AdS/CFT correspondence, the divergence coming from the 
infinite volume of AdS corresponds to the UV divergence in the 
CFT side. The counterterm which cancells the leading divergence in 
the AdS side corresponds to the above action $S_1^{\rm NOO}$.
 
In the present framework, the spacetime inside the brane has  
finite volume and there might be ambiguities when choosing the 
counterterm. Here we give $S_1$ in (\ref{Stotalb}) in terms of 
the superpotential $W(\phi)$ corresponding to (\ref{SuW0}), 
which is given by 
\be
\label{SuW}
V(\phi)={3 \over 4}\left({3 \over 2}\left(
{dW(\phi) \over d\phi}\right)^2 - W(\phi)^2\right)\ .
\ee
This is natural in terms of supersymmetric extension \cite{CLP} 
of the Randall-Sundrum model \cite{RS1,RS2}. This action tells 
that the brane is BPS saturated state and the half of the 
supersymmetries could be conserved \cite{BD}. The factor $2$ 
in front of $S_1$ in (\ref{Stotalb}) is coming from that we 
have two bulk regions which are connected with each other 
by the brane. 

In (\ref{Stotalb}), one chooses 
the 4 dimensional boundary metric as in (\ref{tildeg}).
We should distinguish $A$ and $\tilde g_{\mu\nu}$ with 
$\tilde A(z)$ and $\hat g_{ij}$ in (\ref{AdS}).  
 We also specify the 
quantities given by $\tilde g_{\mu\nu}$ by using $\tilde{\ }$. 

Let us start the consideration of field equations. 
It is often convienient that one assumes 
the metric of 5 dimensional spacetime as follows:
\be
\label{DP1}
ds^2=g_{(5)\mu\nu}dx^\mu dx^\nu =f(y)dy^2 
+ y\sum_{i,j=1}^4\hat g_{ij}(x^k)dx^i dx^j. 
\ee
Here $\hat g_{ij}$ is the metric of 4 dimensional Einstein 
manifold as in (\ref{Ein}). A coordinate corresponding to $z$ 
in (\ref{AdS}) can be obtained by $z=\int dy\sqrt{f(y)}$, 
and solves $y$ with respect to $z$. Then the warp
factor is $\e^{2\hat A(z,k)}=y(z)$. 

On the brane, we obtain the following equations: 
\bea
\label{eq2b}
0&=&{48 l^4 \over 16\pi G}\left(\partial_z A 
 - {1 \over 2}W(\phi)\right)\e^{4A}
+b'\left(4\partial_\sigma^4 A - 16 \partial_\sigma^2 A\right) \nn
&& - 4(b+b')\left(\partial_\sigma^4 A + 2 \partial_\sigma^2 A 
 - 6 (\partial_\sigma A)^2\partial_\sigma^2 A \right) 
+ 2C\left(\partial_\sigma^4 \phi
 - 4 \partial_\sigma^2 \phi \right), \\
\label{eq2pb}
0&=&-{l^4 \over 8\pi G}\e^{4A}\partial_z\phi
 -{3l^3\e^{4A} \over 8\pi G}{dW(\phi) \over d\phi}
+ C\left\{A\left(\partial_\sigma^4 \phi
 - 4 \partial_\sigma^2 \phi \right) 
+ \partial_\sigma^4 (A\phi)
 - 4 \partial_\sigma^2 (A\phi) \right\}\ .
\eea
In (\ref{eq2b}) and (\ref{eq2pb}), using the coordinate $z$ 
and choosing $l^2\e^{2\hat A(z,k)}=y(z)$ one 
uses the form of the metric as in (\ref{metric1}). 
Then 
\bea
\label{smetric}
&& A(z,\sigma)=\hat A(z,k=3) - \ln\cosh\sigma\ ,\quad
A(z,\sigma)=\hat A(z,k=0) + \sigma\ ,\nn
&& A(z,\sigma)=\hat A(z,k=-3) - \ln\sinh\sigma\ ,
\eea
for the unit sphere ($k=3$), 
for the flat Euclidean space ($k=0$), and 
for the unit hyperboloid ($k=-3$), respectively.
We now identify $A$ and $\tilde g$ in (\ref{metric1}) with those in 
(\ref{tildeg}). Then one finds $\tilde F=\tilde G=0$, 
$\tilde R={6 \over l^2}$ etc. 
Note that the sphere corresponds to de Sitter 
space and the hyperboloid to anti-de Sitter 
space when we Wick-rotate the Euclidean signature 
to the Lorentzian one.

By using the equations of motions in the bulk given by 
the Einstein action (\ref{Stotalb}), one can obtain an equation 
that contains only the 
dilaton field $\phi$ (and, of course, bulk potential):
\bea
\label{DP4}
0&=&\left\{ {5k \over 2} - {k \over 4}y^2
\left({d\phi \over dy}\right)^2 + \left({3 \over 2}y 
 + {y^3 \over 6}\left({d\phi \over dy}\right)^2 \right) 
{V(\phi) \over 2}\right\} {d\phi \over dy} \nn
&& + {y^2 \over 2}\left({2k \over y} 
 - {1 \over 2}V(\phi)\right){d^2\phi \over dy^2}
 - \left({3 \over 4} - {y^2 \over 8}
\left({d\phi \over dy}\right)^2 \right){dV(\phi) \over d\phi}\ .
\eea
Several solutions have been found in third ref.\cite{NOO} by 
assuming the dilaton and bulk potentials as:
\be
\label{assmp1}
\phi(y)=p_1\ln \left(p_2 y\right) \ ,\quad 
-V(\phi)= c_1 \exp\left(a\phi\right) 
+ c_2\exp\left(2a\phi\right)\ , 
\ee
where $a$, $p_1$, $p_2$, $c_1$, $c_2$ are some constants. 

When $k\neq 0$, a special solution is given by
\be
\label{case1}
c_1={6kp_2p_1^2 \over 3 - 2p_1^2}\ ,\quad 
c_2=0\ ,\quad a=-{1 \over p_1}\ ,\quad p_1\neq \pm \sqrt{6} \ , 
\quad f(y)={3- 2p_1^2 \over 4ky} \ . 
\ee
Here the superpotential $W(\phi)$ is given by
\be
\label{sW2}
W(\phi)=8p_1^2\sqrt{p_2 k \over (3-2p_1^2)(8p_1^2 - 3)}
\e^{-{\phi \over 2p_1}}\ .
\ee
The potential (\ref{assmp1}) with the coefficients $c_1$ and $c_2$ 
in (\ref{case1}) corresponds to special RG flow in 5d ${\cal N}=8$ 
gauged supergravity where only one scalar from 42 scalars is 
considered. If we define $q^2$ by
\be
\label{q}
q^2\equiv {4k \over 3-2p_1^2}>0\ ,
\ee
the solution when $k=0$ can be obtained by taking $k\rightarrow 0$ 
limit and keeping $q^2$ finite. In the limit, Eqs.(\ref{case1}) 
and (\ref{sW2}) have the following forms:
\bea
\label{case1b}
&& c_1={9 \over 4}q^2 p_2\ ,\quad 
c_2=0\ ,\quad a=-{1 \over p_1}\ ,\quad p_1\neq \pm \sqrt{6} \ ,
\quad f(y)={1 \over q^2 y} \\ 
\label{sW2b}
&& W(\phi)=2\sqrt{q^2 p_2} \e^{-\phi\sqrt{3 \over 2}}\ .
\eea
This solution satisfies Eq.(\ref{Iii}), which is the BPS 
condition, i.e., the solution preserves the half of the 
supersymmetries in the bulk space.  

The solutions in (\ref{case1}) and (\ref{case1b}) have 
a singularity at $y=0$. In fact  
the scalar curvature $R_{(5)}$ is given 
$R_{(5)} = - {3 \over 2}{p_1^2 q^2 \over y}$. 
Here we assume $q^2$ is defined by (\ref{q}) even if $k\neq 0$. 
When $k=3$, the brane becomes de Sitter space after the 
Wick-rotation. Then $y=0$ corresponds to the horizon in the bulk 
5d space. Therefore in $k=3$ case, the singularity is not exactly 
naked. 

In the coordinate system (\ref{DP1}), brane  
Eq.(\ref{eq2pb}) has the following form:
\be
\label{eq2pc}
0=- {y_0^2 \over 8\pi G \sqrt{f(y_0)}}\partial_y\phi 
 - {3y_0^2 \over 8\pi G}{dW(\phi_0) \over d\phi}
 + 6C\phi_0\ .
\ee
Here $\phi_0$ ($\tilde\phi_0$) is the value of the dilaton $\phi$ 
on the brane. Eq.(\ref{eq2b}) has the following form:
\bea
\label{eq2c1}
&0= {3y_0^2 \over 16\pi G}\left({1 \over 2y_0\sqrt{f(y_0)}}
 - {l \over 2}W(\phi_0)\right) + 8b' \quad 
&\mbox{for $k\neq 0$ case} \\
&0= {3y_0^2 \over 16\pi G}\left({1 \over 2y_0\sqrt{f(y_0)}}
 - {l \over 2}W(\phi_0)\right) \quad
&\mbox{for $k=0$ case.} 
\eea
When $k=0$, where $p_1^2={3 \over 2}$, the second equation 
in (\ref{eq2c1}) is satisfied trivially but the first 
equation has the following form:
\be
\label{eq2pc2-0}
-{qy_0^{3 \over 2} \over 8\pi G}
\sqrt{3 \over 2}= 6C\phi_0 \ .
\ee
Then the value $\phi_0$ of the dilaton on the brane depends on $y_0$. 
We should note that the obtained solution for $k=0$ is really 
supersymmetric in the whole system since the corresponding bulk 
solution (\ref{case1b}) satisfies the BPS condition 
Eq.(\ref{Iii}) which tells the solution preserves the half of the 
supersymmetries in the bulk space and the brane action has been 
chosen not to break the supersymmetry on the brane. It is 
interesting that even in case of $k=0$, the quantum effect is 
included in (\ref{eq2pc2-0}) through the parameter $C$ (coefficient 
of dilatonic term in conformal anomaly). In the classical case, 
where $C=0$, the value of the scalar field $\phi_0$ is a free 
parameter. Quantum effects suggest the way for dynamical 
determination of brane dilaton.

When $k\neq 0$, by substituting the solution in (\ref{case1}) 
into (\ref{eq2pc}) and (\ref{eq2c1}), one finds
\bea
\label{eq2pc2}
{p_1y_0^{3 \over 2} \over 4\pi G}
\sqrt{k \over 3-2p_1^2}\left(1 
 - {6 \over \sqrt{8p_1^2 - 3}}\right) &=& 6C\phi_0 \\
\label{eq2c3}
{3 y_0^{3 \over 2} \over 16\pi G}
\sqrt{k \over 3-2p_1^2}\left(1 
 - {2p_1^2 \over \sqrt{8p_1^2 - 3}}\right) &=& 
-8b' \ .
\eea
Since $b'<0$, Eqs.(\ref{eq2pc2}) and (\ref{eq2c3}) have 
non-trivial solutions for $\phi_0$ and $y_0$ if 
\be
\label{cond1}
{k \over 3-2p_1^2}>0
\quad \mbox{and}\quad 
{1 \over 2}<p_1^2<{3 \over 2}\ .
\ee
When $k=3$, where the brane is 4d sphere (de Sitter space when 
we Wick-rotate the brane metric to Lorentzian signature), we have
\be
\label{conddS}
{3 \over 2}>p_1^2 > {7 \over 8}\ .
\ee
On the other hand, if $k=-3$, where the brane is 4d hyperboloid 
(anti-de Sitter after the Wick-rotation), there is no solution 
since the conditions in (\ref{cond1}) conflict with each other. 
Note that de Sitter brane ($k>0$) solution 
does not exist on the classical level but the solution appeared 
after inclusion of the quantum effects of brane matter 
in accordance with AdS/CFT. 

If Eq.(\ref{conddS}) is satisfied, Eqs.(\ref{eq2pc2}) and 
(\ref{eq2c3}) can be explicitly solved with respect to 
$y_0$ and $\phi_0$. This situation is very different from 
 non-supersymmetric case in \cite{NOO}, where $S_1$ was 
chosen as in (\ref{S1noo}). In third ref. from \cite{NOO}, it 
was very difficult to solve the equations corresponding to 
(\ref{eq2pc2}) and (\ref{eq2c3}), explicitly. This indicates 
that supersymmetry simplifies the situation and the approach 
we adopt is right way to construct supersymmetric new brane world. 
Moreover, quantum effects may give a natural mechanism for 
SUSY breaking.
 
If one writes $y_0=R_b^2$, $R_b$ corresponds to the radius of the 
sphere ($k=3$). Since $b'\propto N^2$ for large $N$, 
from Eqs.(\ref{AdSCFT}) and 
(\ref{eq2c3}), one gets 
\be
\label{RN}
R_b\propto \left(GN^2\right)^{1 \over 3}\sim N^{3 \over 4}\ .
\ee

Despite some modern progress, it is clear that much work is necessary
in order to construct consistent supersymmetric New Brane World.

\section{Gravity perturbations}

It is known that brane gravity trapping occurs on curved brane 
in a different way than on flat brane. For example, in 
refs.\cite{KR,KMP} 
 the AdS$_4$ branes in AdS$_5$ were discussed 
and the existence of the massive normalizable mode of graviton 
was found. In these papers, the tensions of the branes are free 
parameters but in the case treated in the present section the 
tension is dynamically determined. As brane solutions are 
found in the previous section when the brane is flat or de Sitter 
space, it is reasonable to consider  perturbation around 
the solution. 

Let us regard the brane as an object with a tension $\tilde U(\phi)$ 
and assume the brane can be effectively described by the 
following action, as in (\ref{S}) (for simplicity, we only 
consider the brane corresponding to $i=2$, or the limit 
$z_1\rightarrow -\infty$):
\be
\label{bten1}
S_{\rm brane}=- {1 \over 16\pi G}\int d^4x \sqrt{-g_{(4)}}
\tilde U(\phi)\ .
\ee
Then using the Einstein equation as in (\ref{Eiii}), one finds 
\be
\label{bten2}
A'' + 4 (A')^2 = -{V \over 3}
 - {\tilde U(\phi) \over 6}\delta(z-z_0) \ .
\ee
Here we assume that there is a brane at $z=z_0(=z_2)$. 
Then at $z=z_0$
\be
\label{bten3}
\left. A' \right|_{z=z_0}=-{\tilde U(\phi) \over 6}\ .
\ee
Comparing (\ref{bten3}) 
with (\ref{eq2b}) etc. one gets, when $k\neq 0$ 
\be
\label{bten4}
\tilde U(\phi) = - {3 \over l}W(\phi_0) 
+ {48\pi G b' \over R_b^4}\ .
\ee
and when $k=0$
\be
\label{bten4b}
\tilde U(\phi) = - {3 \over l}W(\phi_0) \ .
\ee
Note that tension becomes $R_b$ dependent due to the quantum 
correction when $k\neq 0$, as $b'\sim {\cal O}\left(N^2\right)$ 
and $R_b^4\sim {\cal O}\left(N^3\right)$ from 
(\ref{RN}), the tension depends on $N$ as 
$\tilde U(\phi) + {3 \over l}W(\phi_0) 
\sim {\cal O}\left(N^{-{9 \over 4}}\right)$. 
One can understand that the r.h.s. in (\ref{bten4b}) and the 
first term in the r.h.s. in (\ref{bten4}) are determined 
from the supersymmetry.

 Consider the perturbation by assuming the metric 
in the following form:
\be
\label{bten5}
ds^2=\e^{2\hat A(\zeta)}\left(d\zeta^2 + \left(\hat g_{\mu\nu}
+ \e^{-{3 \over 2}\hat A(\zeta)}h_{\mu\nu}\right) dx^\mu dx^\nu\right)\ .
\ee
Here the gauge conditions $h^\mu_{\ \mu}=0$ and 
$\nabla^\mu h_{\mu\nu}=0$ are chosen. 
Then one obtains the following equation
\be
\label{bten7}
\left(-\partial_\zeta^2 + {9\over 4} \left(\partial_\zeta \hat A
\right)^2 + {3 \over 2}\partial_\zeta^2 \hat A \right)h_{\mu\nu}
= m^2 h_{\mu\nu}
\ee
Here $m^2$ corresponds to the mass of the graviton on the brane 
\be
\label{bten8}
\left(\htBox + {1 \over R_b^2}\right)h_{\mu\nu} = m^2 
h_{\mu\nu}\ .
\ee
for $k>0$ and 
\be
\label{bten8b}
\htBox h_{\mu \nu} = m^2 
h_{\mu\nu}
\ee
for $k=0$. Here $\htBox$ is 4 dimensional d'Alembertian 
constructed on $\hat g_{\mu\nu}$. Since
\be
\label{bten9}
\pm\e^{A}d\zeta = dz 
= \sqrt{f}dy\ ,\quad \e^A={\sqrt{y} \over l}\ ,
\ee
one finds
\be
\label{bten10}
\pm\zeta=\int dy \sqrt{f(y) \over y} \ .
\ee
If we choose $\zeta=0$ when $y=y_0$, Eq.(\ref{bten10}) 
for the solution in (\ref{case1}) or (\ref{case1b}) gives 
\be
\label{bten12}
|\zeta|=-{1 \over q}\ln y +{1 \over q}\ln y_0\ .
\ee
Here we assume $q$ is defined by (\ref{q}) even if 
$k>0$. Since only the square of $q$ is defined in (\ref{q}), we 
can choose $q$ to be positive. 

Note that brane separates two bulk regions corresponding to 
$\zeta<0$ and $\zeta>0$, respectively.
Since $y$ takes the value in $[0,y_0]$, $\zeta$ takes the 
value in $[-\infty,\infty]$. 
Since $A={1 \over 2}\ln y$, from (\ref{bten7}) one gets
\be
\label{bten13}
\left(-\partial_\zeta^2 + {9q^2 \over 4} - 3q\delta(\zeta)
\right)h_{\mu\nu} = m^2 h_{\mu\nu}
\ee
Zero mode solution with $m^2$ of (\ref{bten13}) is given by
\be
\label{bten14}
h_{\mu\nu}=h^{(0)}_{\mu\nu}\e^{-{3q \over 2}|\zeta|}\ .
\ee
Here $h^{(0)}_{\mu\nu}$ is a constant. Any other normalizable 
solution does not exist. When 
\be
\label{min}
m^2>{9 \over 4}q^2\ ,
\ee 
there are non-normalizable solutions given by
\be
\label{bten15}
h_{\mu\nu}=a_{\mu\nu}
\cos \left(|\zeta|\sqrt{m^2-{9 \over 4}q^2}\right) 
+ b_{\mu\nu} \sin  \left(|\zeta|
\sqrt{m^2-{9 \over 4}q^2}\right) \ .
\ee
The coefficients $a_{\mu\nu}$ and $b_{\mu\nu}$ are constants of 
the integration and they are determined to satisfy the boundary 
condition, which comes from the $\delta$-function potential 
in (\ref{bten13}), 
\be
\label{bten16}
\left.{\partial_\zeta h_{\mu\nu} \over h_{\mu\nu}}
\right|_{\zeta\rightarrow 0+}
=-{3 \over 2}q\ .
\ee
Note that zero mode solution (\ref{bten14}) satisfies this 
boundary condition (\ref{bten16}). By using (\ref{bten16}), we 
can determine the coefficients $a_{\mu\nu}$ and 
$b_{\mu\nu}$ for non-normalizable solutions as follows:
\be
\label{ab}
a_{\mu\nu}=-{2b_{\mu\nu} \over 3q}\sqrt{m^2-{9 \over 4}q^2}\ .
\ee
It might be interesting that there is the minimum (\ref{min}) 
in the mass of non-normalizable mode. This situation is different 
from the original Randall-Sundrum model \cite{RS2}. Although de 
Sitter brane appears when we include the quantum correction, the 
minimum itself does not depend on the parameter of the 
quantum correction $b'$ or $N$. 

Since there is only one normalizable solution corresponding 
to zero mode (\ref{bten14}) and other solutions (\ref{bten15}) are 
non-normalizable,  gravity should be localized on the brane 
and the leading long range potential between two massive sources 
on the brane should obey the Coulomb law, i.e., 
${\cal O}\left(r^{-1}\right)$. Here $r$ is the distance between 
the above two massive sources. 
Furthermore the existence of the the minimum (\ref{min}) 
in the mass of non-normalizable mode indicates that the 
correction to the Coulomb law should be small. 

\section{AdS/CFT and quantum-corrected brane entropy}

In the previous sections, we have only considered the case that 
the radius of the brane 
 is constant. 
In this section,  the situation that radius depends 
on the ``time'' coordinate is discussed. If we consider the AdS-Schwarzschild 
background, the obtained equation, which 
describes the dynamics of the brane, can be regarded as the 
Friedmann-Robertson-Walker (FRW) equation. 
>From the equation,  one obtains the quantum-corrected brane 
entropy as well as Hubble constant and Hawking temperature. 

Let us start with the Minkowski 
signature action. Then  being in the Brane New World, we have the following 
equation which generalizes the 
classical brane equation of the motion:
\bea
\label{eq2bb}
0&=&{48 l^4 \over 16\pi G}\left(A_{,z} 
 - {1 \over l}\right)\e^{4A}
+b'\left(4 \partial_\tau^4 A + 16 \partial_\tau^2 A\right) \nn
&& - 4(b+b')\left(\partial_\tau^4 A - 2 \partial_\tau^2 A 
 - 6 (\partial_\tau A)^2\partial_\tau^2 A \right) \ .
\eea
In (\ref{eq2bb}), one 
uses the form of the metric as 
\be
\label{metric1b}
ds^2=dz^2 + \e^{2A(z,\tau)}\tilde g_{\mu\nu}dx^\mu dx^\nu\ ,
\quad \tilde g_{\mu\nu}dx^\mu dx^\nu\equiv l^2\left(-d \tau^2 
+ d\Omega^2_3\right)\ .
\ee
Here $d\Omega^2_3$ corresponds to the metric of 3 dimensional 
unit sphere. 

As a bulk space,  one takes 5d AdS-Schwarzschild space-time, 
whose metric is given by
\be
\label{AdSS} 
ds_{\rm AdS-S}^2 = {1 \over h(a)}da^2 - h(a)dt^2 
+ a^2 d\Omega_3^2 \ ,\ \ 
h(a)= {a^2 \over l^2} + 1 - {16\pi GM \over 3 V_3 a^2}\ .
\ee
Here $V_3$ is the volume of the unit 3 sphere.  
If one chooses new coordinates $(z,\tau)$ by
\be
\label{cc1}
{\e^{2A} \over h(a)}A_{,z}^2 - h(a) t_{,z}^2 = 1 \ ,
\quad {\e^{2A} \over h(a)}A_{,z}A_{,\tau} - h(a)t_{,z} t_{,\tau}
= 0 \ ,\quad 
{\e^{2A} \over h(a)}A_{,\tau}^2 - h(a) t_{,\tau}^2 
= -\e^{2A}\ .
\ee
the metric takes the form (\ref{metric1b}). Here $a=l\e^A$.
Furthermore choosing a coordinate $\tilde t$ by 
$d\tilde t = l\e^A d\tau$, 
the metric on the brane takes FRW form: 
\be
\label{e3}
\e^{2A}\tilde g_{\mu\nu}dx^\mu dx^\nu= -d \tilde t^2  
+ l^2\e^{2A} d\Omega^2_3\ .
\ee
By solving Eqs.(\ref{cc1}), we have
\be
\label{e4}
H^2 = A_{,z}^2 - h\e^{-2A}= A_{,z}^2 - {1 \over l^2}
 - {1 \over a^2} + {16\pi GM \over 3 V_3 a^4}\ .
\ee
Here the Hubble constant $H$ is defined by
$H={dA \over d\tilde t}$. Finally, after some algebra one arrives to
\bea
\label{e10}
H^2 &=& - {1 \over a^2} 
+ {8\pi G_4 \rho \over 3} \\
\label{e10b}
\rho&=&{l \over a}\left[ {M \over V_3 a^3} \right. 
+ {3a \over 16\pi G}\left[
\left[{1 \over l} + {\pi G \over 3}\left\{ 
-4b'\left(\left(H_{,\tilde t \tilde t \tilde t} + 4 H_{,\tilde t}^2 
+ 7 H H_{,\tilde t\tilde t} \right.\right.\right.\right. \right.\nn
&& \left.\left. + 18 H^2 H_{,\tilde t} + 6 H^4\right) 
+ {4 \over a^2} \left(H_{,\tilde t} + H^2\right)\right) 
+ 4(b+b') \left(\left(H_{,\tilde t \tilde t \tilde t} 
+ 4 H_{,\tilde t}^2 \right. \right. \nn
&& \left.\left.\left.\left.\left.\left. + 7 H H_{,\tilde t\tilde t} 
+ 12 H^2 H_{,\tilde t} \right) - {2 \over a^2} 
\left(H_{,\tilde t} + H^2\right)\right) \right\}\right]^2
 - {1 \over l^2} \right]\right]\ .
\eea
This can be regarded as the quantum FRW equation of the brane universe.
Here 4d Newton constant $G_4$ is given by
\be
\label{e12}
G_4={2G \over l}\ .
\ee
Eq.(\ref{e10}) expresses the quantum correction to the corresponding 
equation in \cite{SV}. In fact, if we put $b=b'=0$, 
Eq.(\ref{e10}) reduces to the classical one. 
Further by differentiating Eq.(\ref{e10}) with respect to 
$\tilde t$, we obtain the second FRW equation
\bea
\label{e11}
H_{,\tilde t} &=&  {1 \over a^2} - 4\pi G_4(\rho + p) \\
\label{e11b}
\rho + p &=& {l \over a}\left[
 {4 M \over 3 V_3 a^3} \right. - {1 \over 24l^3 H}\left[{1 \over l} 
+ {\pi G \over 3}\left\{ 
-4b'\left(\left(H_{,\tilde t \tilde t \tilde t} + 4 H_{,\tilde t}^2 
+ 7 H H_{,\tilde t\tilde t} \right.\right.\right.\right. \nn
&& \left.\left. + 18 H^2 H_{,\tilde t} + 6 H^4\right) 
+ {4 \over a^2} \left(H_{,\tilde t} + H^2
\right)\right) \nn
&& + 4(b+b') \left(\left(H_{,\tilde t \tilde t \tilde t} 
+ 4 H_{,\tilde t}^2 + 7 H H_{,\tilde t\tilde t} 
+ 12 H^2 H_{,\tilde t} \right) 
\left.\left. - {2 \over a^2} \left(H_{,\tilde t} + H^2\right) 
\right)\right\}\right] \nn
&& \times \left\{ 
-4b'\left(\left(H_{,\tilde t \tilde t \tilde t \tilde t} 
+ 15 H_{,\tilde t} H_{\tilde t\tilde t} 
+ 7 H H_{,\tilde t\tilde t\tilde t} 
+ 18 H^2 H_{,\tilde t\tilde t} 
+ 36 H H_{,\tilde t}^2 \right.\right.\right. \nn
&& \left.\left. + 24 H^3 H_{,\tilde t} \right)
+ {4 \over a^2} \left(H_{,\tilde t\tilde t} - 2 H^3\right) \right) 
+ 4(b+b') \left(\left(H_{,\tilde t \tilde t \tilde t \tilde t} 
+ 15 H_{,\tilde t} H_{,\tilde t\tilde t} \right.\right. \nn
&& \left.\left. 
+ 7 H H_{,\tilde t\tilde t\tilde t} + 12 H^2 H_{,\tilde t\tilde t}
+ 24 H H_{,\tilde t}^2 \right)
\left.\left. - {2 \over a^2} \left(H_{,\tilde t\tilde t} 
 - 2H^2\right)\right) \right\}\right]\ .
\eea
The quantum corrections from CFT are included into the definition of
energy (pressure). These quantum corrected FRW equations are written
from quantum-induced brane-world perspective. Similar equations 
from the point of view of 4d brane observer (who does not 
know about 5d AdS bulk) have been presented in ref.\cite{NO}.
Clearly, brane-world approach gives more information.

Note that when $a$ is large, the metric 
(\ref{AdSS}) has the following form:
\be
\label{eq13} 
ds_{\rm AdS-S}^2 \rightarrow {a^2 \over l^2}\left(dt^2 
+ l^2 d\Omega_3^2\right)\ ,
\ee
which tells that the CFT time $t_{\rm CFT}$ is equal 
to the AdS time $t$ times the factor ${a \over l}$: 
$t_{\rm CFT}={a \over l}t$.
Therefore the energy $E_{\rm CFT}$ in CFT is related 
with the energy $E_{\rm AdS}$ in AdS by \cite{SV}
\be
\label{eq15}
E_{\rm CFT}={l \over a}E_{\rm AdS}\ .
\ee
The factor ${l \over a}$ in front of Eqs.(\ref{e10b}) and 
(\ref{e11b}) appears due to the above scaling of the energy 
in (\ref{eq15}) or time. 

The AdS$_5$-Schwarzschild black hole solution in (\ref{AdSS}) 
has a horizon at $a=a_H$, where $h(a)$ vanishes \cite{HP}:
\be
\label{eq16}
h(a_H)= {a_H^2 \over l^2} + 1 - {16\pi GM \over 3 V_3 a_H^2}=0\ .
\ee
Then considering the moment the brane crosses these points and
using (\ref{e10}),  one gets
\bea
\label{e17}
H &=& \pm \left[{1 \over l} + {\pi G \over 3}\left\{ 
-4b'\left(\left(H_{,\tilde t \tilde t \tilde t} + 4 H_{,\tilde t}^2 
+ 7 H H_{,\tilde t\tilde t} + 18 H^2 H_{,\tilde t} + 6 H^4\right) 
+ {4 \over a_H^2} \left(H_{,\tilde t} 
+ H^2\right)\right) \right.\right. \nn
&&\left.\left.+ 4(b+b') \left(\left(H_{,\tilde t \tilde t \tilde t} 
+ 4 H_{,\tilde t}^2 + 7 H H_{,\tilde t\tilde t} 
+ 12 H^2 H_{,\tilde t} \right) - {2 \over a_H^2} \left(H_{,\tilde t} 
+ H^2\right)\right) \right\}\right]\ .
\eea
The sign $\pm$ depends on whether the brane is expanding or 
contracting. 
Obviously, if the higher derivative of the Hubble constant $H$ is 
large, the quantum correction becomes essential. 

We now assume that the brane behaves as de Sitter 
(inflationary) space $a=A\cosh B\tilde t$ 
near the horizon. This is quantum-induced 
brane universe. 
Note that this is not the solution for positive 
(non-vanishing) black hole mass $M>0$ but the above assumption 
is very natural. Then Eqs.(\ref{eq16}) and (\ref{e17}) 
have the following forms: 
\bea
\label{eq16b}
h(a_H) &=& {A^2\cosh^2 \tilde t_H \over l^2} + 1
 - {16\pi GM \over 3 V_3 A^2\cosh^2 \tilde t_H }=0 \\
\label{e17b}
H &=& \pm \left[{1 \over l} + {\pi G \over 3}\left\{ 
-4b'\left(-4\left(B^4 - {B^2 \over A^2}\right)
{1 \over \cosh^2 \tilde t_H} + 6B^4\right) \right.\right. \nn
&& \left. + 8(b + b') \left(B^4 - {B^2 \over A^2}\right)
{1 \over \cosh^2 \tilde t_H}\right]\ .
\eea
Here the brane crosses the horizon when $\tilde t=\tilde t_H$. 
Thus, quantum-corrected Hubble parameter at the horizon is defined.
The quantum correction becomes large when the rate $B$ of expansion 
of the universe is large. 

Let the entropy ${\cal S}$ of CFT on the brane is given 
by the Bekenstein-Hawking entropy of the AdS$_5$ black hole 
${\cal S}={V_H \over 4G}$. 
Here $V_H$ is the area of the horizon, which is equal to the 
spatial brane when the brane crosses the horizon:
$V_H=a_H^3 V_3$. 
If the total entropy ${\cal S}$ is constant during the 
cosmological evolution, the entropy density $s$ is given 
by (see \cite{SV})
\be
\label{e20}
s={{\cal S} \over a^3 V_3}= {l a_H^3 \over 2G_4 a^3}\ .
\ee
Here  Eq.(\ref{e12}) is used. The expression in (\ref{e20}) 
is identical with the classical one. The quantum correction 
appear when we express $s$ in terms of the quantities in 
brane universe, say $H$, $H_{,\tilde t}$ etc., by using 
(\ref{e17}). 

The Hawking temperature of the black hole is given by (see \cite{SV}) 
\be
\label{e21}
T_H={h'(a_H) \over 4\pi} 
={a_H \over 2\pi l^2} + {8GM \over 3V_3 a_H^3} 
={a_H \over \pi l^2} + {1 \over 2\pi a_H}\ .
\ee
Here (\ref{eq16}) is used. 
As in (\ref{eq15}), the temperature $T$ on the brane 
is different from that of AdS$_5$ by the factor 
${l \over a}$:
\be
\label{e22}
T={l \over a}T_H
={a_H \over \pi a l} + {l \over 2\pi a a_H}\ .
\ee
Especially when $a=a_H$
\be
\label{e23}
T={1 \over \pi l} + {l \over 2\pi a_H^2}\ .
\ee
For the solution in the form of 
$a=A\cosh B\tilde t$, one gets
\be
\label{e9bb}
T = {l\over \pi}\left[ -
H_{,\tilde t}  \pm {\pi G \over 3 }(48b' + 16b) 
\left(B^4 - {B^2 \over A^2}\right) {\sinh B\tilde t_H 
\over \cosh^3 B\tilde t_H}\right]\ .
\ee
The quantum correction becomes dominant when $B\tilde t_H$ 
is of order 
unity but $B$ ($\neq {1 \over A}$) is large or $A$ is small. 
Since the radius of the horizon is given by 
$a_H=A\cosh B\tilde t_H$, this might mean that if quantum 
correction is large then the radius of the black hole 
is small.

In \cite{EV} it has been pointed out that the first FRW equation 
(\ref{e10}) can be regarded as the $4$-dimensional analogue 
of the Cardy formula \cite{Cardy} for the entropy 
${\cal S}$ of the CFT on the brane. In fact identifying
\be
\label{CV2}
{2\pi \rho V a \over 3} \Rightarrow 2\pi L_0 \ ,\quad 
{V \over 8\pi G_4 a} \Rightarrow {c \over 24} \ ,\quad 
{4\pi HV \over 8\pi G_4} \Rightarrow {\cal S}_H\ ,
\ee
 one finds
\be
\label{CV1}
{\cal S}=2\pi \sqrt{
{c \over 6}\left(L_0 - {c \over 24}\right)}\ .
\ee
Furhtermore if we define the Bekenstein entropy $S_B$ 
and the Bekenstein-Hawking entropy $S_{BH}$ by
\be
\label{CV7}
S_B={2\pi \over 3}Ea=2\pi L_0\ ,
\quad S_{BH}={\pi V \over 2G_4 a}={\pi \over 6}c\ ,
\ee
one gets 
\be
\label{CV8}
{\cal S}^2 = 2S_B S_{BH} - S_{BH}^2 \ ,
\ee
or
\be
\label{CV8b}
{\cal S}^2 + \left(S_{BH} - S_B\right)^2 = S_B^2\ ,
\ee
which gives the dynamical bound for  cosmological entropies. It is quite 
possible that the origin of cosmological entropy bounds in quantum gravity
 should be searched in this direction.

\section{Summary}

In summary, we reviewed the New Brane World  
\cite{HHR,NOZ,NOplb} and the attempt to 
supersymmetrize the quantum-induced 
dilatonic New Brane World \cite{NOOs}. 
Furthermore it was shown \cite{NOent} 
that inside d5 AdS BH the inflationary brane induced by CFT 
quantum effects in accordance with AdS/CFT may occur in 
the same way as in refs.\cite{HHR, NOZ,NOplb}.

It is 
shown that for a number of superpotentials one can construct 
flat SUSY dilatonic brane-world or de Sitter dilatonic brane-world 
where SUSY is broken by quantum effects. The crucial role in the 
creation of de Sitter brane universe is in account of quantum 
effects which produce the effective brane 
tension. The analysis of graviton perturbations for such 
brane-worlds shows that gravity trapping on the brane occurs.

The stress tensor of the inflationary brane inside d5 AdS BH 
is completely defined by dual quantum CFT (and also probably, 
by brane QG) and it is not chosen by hands as it  happens often 
in the traditional brane-world scenarios.   
The similirity between CFT entropy at the horizon and FRW equations 
discovered in refs.\cite{EV,SV} is extended for 
the presence of quantum effects in Brane New World.
 Such approach may be important for the generalization of 
cosmological entropy bounds in the case of quantum gravity. From 
another side, it would be interesting to use such study with the
purpose of extension of AdS/CFT correspondence for cosmological 
(AdS) backgrounds \cite{NOA}.

To conclude, New Brane World represents the embedding into brane physics 
of old known trace-anomaly driven inflationary scenario (for recent review, see \cite{HHR2}). Its relation with AdS/CFT correspondence, possibility to
extend it to presence of dilaton or higher derivative bulk terms \cite{HDSO} or
supersymmetrize it, natural connection with cosmological entropy bounds
indicates to its important role in the construction of realistic theory 
of early Universe evolution. Last but not least remark is that it 
maybe naturally discussed in frames of M-theory.

\section*{Acknowledgements}

The work of S.D.O. has been supported in part by CONACyT (CP, 
ref.990356 and grant 28454E). S.N. is very indebted to Y. Machida 
for the hospitality in Numazu and also to people in Osaka University, 
especially to H. Itoyama. 

\appendix

\section{Various expressions of AdS}

$D=d+1$-dimensional Euclidean anti de Sitter space can be 
embedded in $D+1$-dimensional flat space, whose metric is 
given by
\be
\label{AA1}
ds_{D+1}^2 = \left(dX^1\right)^2 + \left(dX^2\right)^2 
+ \cdots + \left(dX^D\right)^2 - \left(dX^0\right)^2 \ .
\ee
The AdS space is the hypersurface given by
\be
\label{AA2}
\left(X^1\right)^2 + \left(X^2\right)^2 
+ \cdots + \left(X^D\right)^2 - \left(X^0\right)^2 = - l^2\ .
\ee
Here $l$ is a constant, which gives the length scale of the 
AdS. If we define new coordinates $U$, $V$ and $x^i$ by
\be
\label{AA3}
V=X^D - X^0\ ,\quad U=X^D + X^0\ ,\quad X^i=Ux^i\ (i=1,2,\cdots,
d(=D-1)
\ee
and solve (\ref{AA2}) with respect to $V$ as a function of 
$U$ and $x^i$, we obtain the following metric
\be
\label{A4}
ds_{\rm AdS}^2 = {l^2 \over U^2}dU^2 + U^2 \left\{
\left(dx^1\right)^2 + \left(dx^2\right)^2 + \cdots 
+ \left(dx^1\right)^2 \right\}\ .
\ee
Further if  one changes the variable $U$ by 
$U=\e^{y \over l}$  
\be
\label{A6}
ds_{\rm AdS}^2 = dy^2+ \e^{2y \over l} \left\{
\left(dx^1\right)^2 + \left(dx^2\right)^2 + \cdots 
+ \left(dx^1\right)^2 \right\}\ .
\ee

On the other hand one can choose the polar coordinates for 
$\left(X^1,X^2,\cdots,X^D\right)$ and let $Y$ be a 
radial coordinate
\be
\label{A7}
Y^2=\left(X^1\right)^2 + \left(X^2\right)^2 
+ \cdots + \left(X^D\right)^2 
\ee
Then by deleting $X^0$ by using (\ref{AA2}), 
we obtain
\be
\label{A8}
ds_{\rm AdS}^2 ={l^2dY^2 \over Y^2 + l^2} + Y^2 d\Omega_d^2\ .
\ee
Here $d\Omega_d^2$ is the metric of the $d$-dimensional sphere. 
If we change the variable by $Y=l\sinh {y \over l}$ 
\be
\label{A10}
ds_{\rm AdS}^2 =dy^2 + l\sinh^2 y d\Omega_d^2\ .
\ee

 One can also choose a coordinate $Z$ by
\be
\label{A11}
-Z^2=\left(X^1\right)^2 + \left(X^2\right)^2 
+ \cdots + \left(X^{D-1}\right)^2 - \left(X^0\right)^2 \ .
\ee
Then the hypersurface on the $D$-dimensional AdS with constant 
$Z$ is the $D-1=d$-dimensional AdS. Let the metric of 
$d$-dimensional AdS with unit length parameter is given by $dH_d^2$. 
Then the metric of $D$-dimensional AdS is
\be
\label{A12}
ds_{\rm AdS}^2 ={l^2dZ^2 \over Y^2 - l^2} + Z^2 dH_d^2\ .
\ee
Further changing the variable by  $Z=l\cosh {y \over l}$ one gets
\be
\label{A14}
ds_{\rm AdS}^2 =dy^2 + l\cosh^2 y dH_d^2\ .
\ee

\end{document}